\documentclass[sn-mathphys,Numbered]{sn-jnl}

\usepackage{graphicx}%
\usepackage{multirow}%
\usepackage{amsmath,amssymb,amsfonts}%
\usepackage{amsthm}%
\usepackage{mathrsfs}%
\usepackage[title]{appendix}%
\usepackage{xcolor}%
\usepackage{textcomp}%
\usepackage{manyfoot}%
\usepackage{booktabs}%
\usepackage{algorithm}%
\usepackage{algorithmicx}%
\usepackage{algpseudocode}%
\usepackage{listings}%
\usepackage{rotating}

\usepackage{epsfig}
\usepackage{amsmath}
\usepackage{graphicx}
\usepackage{pifont}
\usepackage{xcolor}
\usepackage{lineno}

\begin{document}

\title[X-ray AGN in a $z\sim10$ galaxy]{Evidence for heavy seed origin of early supermassive black holes from a $z\sim10$ X-ray quasar}


\author*[1]{\fnm{\'Akos} \sur{Bogd\'an}}\email{abogdan@cfa.harvard.edu}

\author[2]{\fnm{Andy D.} \sur{Goulding}}
\equalcont{These authors contributed equally to this work.}

\author[3,4,5]{\fnm{Priyamvada} \sur{Natarajan}}
\equalcont{These authors contributed equally to this work.}

\author[6]{\fnm{Orsolya E.} \sur{Kov\'acs}}

\author[1]{\fnm{Grant R.} \sur{Tremblay}}

\author[1]{\fnm{Urmila} \sur{Chadayammuri}}

\author[7]{\fnm{Marta} \sur{Volonteri}}

\author[1]{\fnm{Ralph P.} \sur{Kraft}}

\author[1]{\fnm{William R.} \sur{Forman}}

\author[1]{\fnm{Christine} \sur{Jones}}

\author[8]{\fnm{Eugene} \sur{Churazov}}

\author[9]{\fnm{Irina} \sur{Zhuravleva}}

\affil*[1]{Center for Astrophysics \ding{120} Harvard \& Smithsonian, 60 Garden Street, Cambridge, MA 02138, USA}

\affil[2]{Department of Astrophysical Sciences, Princeton University, Princeton, NJ 08544, USA}

\affil[3]{Department of Astronomy, Yale University, 52 Hillhouse Avenue, New Haven, CT 06511, USA}

\affil[4]{Department of Physics, Yale University, P.O. Box 208121, New Haven, CT 06520, USA}

\affil[5]{Black Hole Initiative, Harvard University, 20 Garden Street, Cambridge, MA 02138}
\affil[6]{Department of Theoretical Physics and Astrophysics, Faculty of Science, Masaryk University, Brno, 611 37, Czech Republic}
\affil[7]{Institut d'Astrophysique de Paris, Sorbonne Universit\'e, CNRS, UMR 7095, 98 bis bd Arago, 75014 Paris, France}
\affil[8]{Max Planck Institut f\"ur Astrophysik, Karl-Schwarzschild-Str.1, 85741 Garching bei M\"unchen, Germany}
\affil[9]{Department of Astronomy and Astrophysics, The University of Chicago, Chicago, IL 60637, USA }


\abstract{\textbf{Observations of quasars reveal that many supermassive black holes (BHs) were in place less than 700 million years after the Big Bang. However, the origin of the first BHs remains a mystery. Seeds of the first BHs are postulated to be either light (i.e., $\mathbf{10-100~\rm{M_{\odot}})}$, remnants of the first stars or heavy (i.e., $\mathbf{10^4-10^5~\rm{M_{\odot}})}$, originating from the direct collapse of gas clouds. Harnessing recent data from the \textit{Chandra X-ray Observatory}, we report the detection of an X-ray-luminous massive BH in a gravitationally-lensed galaxy identified by {\textit{JWST}} at $\mathbf{z\approx10.3}$ behind the cluster lens Abell~2744. This heavily-obscured quasar with a bolometric luminosity of $\mathbf{L_{\rm bol}\sim5\times10^{45}~\rm{\mathbf{erg\ s^{-1}}}}$ harbors a $\mathbf{M_{\rm BH}\sim10^7-10^8~\rm{M_{\odot}}}$ BH assuming accretion at the Eddington limit. This mass is comparable to the inferred stellar mass of its host galaxy, in contrast to what is found in the local Universe wherein the BH mass is $\sim0.1\%$ of the host galaxy's stellar mass. The combination of such a high BH mass and large BH-to-galaxy stellar mass ratio just $\sim$500~Myrs after the Big Bang was theoretically predicted and is consistent with a picture wherein BHs originated from heavy seeds.}}





\maketitle

Data from \textit{JWST} is rapidly transforming our understanding of the early universe by enabling the detection of large samples of faint, distant galaxies deep into the epoch of reionization. Early studies are hinting at a higher-than-expected abundance of galaxies in the early Universe\cite{castellano22a,castellano22b,harikane22,naidu22,atek23,adams23}. Cluster gravitational lenses, nature's telescopes, further augment \textit{JWST}'s sensitivity by bringing into view low-mass, faint galaxies at the highest redshifts. As part of an Early Release Science Program\cite{treu22} and a Cycle 1 Treasury Program\cite{bezanson22}, \textit{JWST} has peered through the Hubble Frontier Fields cluster lens Abell~2744 at $z=0.308$, revealing a high galaxy density\cite{castellano22a,castellano22b} at $z>9$. In this field, multiple independent methods have ascertained accurate photometric redshifts for several $z\sim (9-15)$ galaxies, and their physical properties, such as their stellar mass and star-formation rate, have been derived from fitting their spectral energy distributions (SEDs)\cite{castellano22a,castellano22b}.

\begin{figure}  
\centering
  \includegraphics[width=0.99\textwidth]{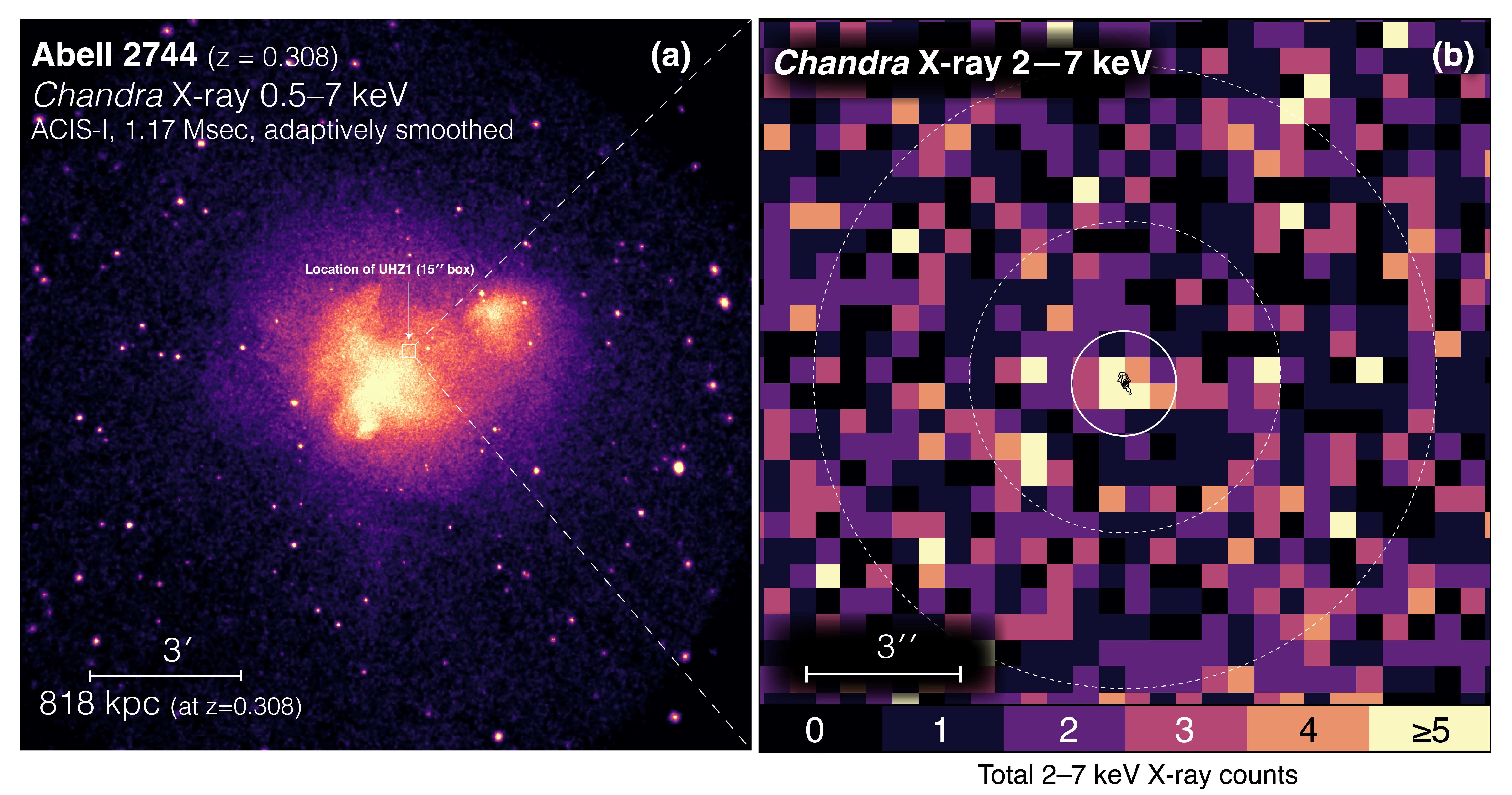}
 \caption{A $4.2\sigma$ \textit{Chandra} X-ray detection of a source cospatial with UHZ1: Panel (a) 1.25~Msec \textit{Chandra} X-ray image of $0.5-7$ keV emission associated with the galaxy cluster lens Abell~2744. The image has been smoothed with an adaptive Gaussian kernel, and a logarithmic stretch has been applied. The $15'' \times 15''$ white box is centered on the location of UHZ1 and shows the zoomed-in field of view of the adjacent panel. Panel (b) \textit{Chandra} $2-7$~keV band X-ray image of the $15'' \times 15''$ region surrounding UHZ1. Black contours show the \textit{JWST}/NIRCam F200W morphology of the UHZ1 galaxy candidate at $z\approx10.32$. The solid white circle has a $1''$ radius, corresponding to both the off-axis \textit{Chandra} PSF ($\approx88\%$ encircled counts fraction) at this location on the image, as well as the source spectral extraction region as described in the text. The dashed white circle has an inner and outer radius of $3''$ and $6''$ respectively, marking the background spectral extraction region. North is up and East is left.}
 \label{fig:chandra}
\end{figure}

To address the fundamental open question about the origin of the first BHs, we deployed an innovative approach to probe galaxies and their central BHs at cosmic dawn, utilizing the achromatic, magnifying power of the extremely well-calibrated cluster lens Abell~2744 with deep \textit{Chandra} X-ray observations. X-rays are a ubiquitous signature of BH accretion, and hard X-ray observations can detect accreting BHs even if they are surrounded by large columns of absorbing gas and dust. To this end, we search for X-ray emission associated with high redshift accreting BHs in \textit{JWST}-detected lensed galaxies in the Abell~2744 field using $1.25$~Ms deep \textit{Chandra} imaging observations. 

The processing and analysis of the \textit{Chandra} X-ray data in Abell~2744 were carried out with \textsc{CIAO}\cite{fruscione06}. We reprocessed the observations using the \textsc{chandra\_repro} tool to apply the latest calibration data. The individual observations were merged and we generated images in the $0.5-7$~keV, $0.5-2$~keV, and $2-7$~keV bands with the \textsc{merge\_obs} task. For each energy range, maps of the point spread function were generated, weighted, and co-added. The absolute astrometry of each observation was corrected using a set of bright X-ray point sources and by employing the \textsc{wcs\_match} and \textsc{wcs\_update} tools before merging the X-ray images. Details on the data analysis are presented in the Methods Section.

To probe the presence of high-redshift X-ray sources magnified by Abell~2744, we investigated a sample of 11 \textit{JWST}-detected, gravitationally-lensed galaxies with high signal-to-noise and $z>9$ photometric redshifts\cite{castellano22a,castellano22b}. Using the \textit{JWST} positions of the $z>9$ galaxies, we performed photometry on the merged \textit{Chandra} X-ray images. To minimize the contribution of the intracluster medium (ICM) of Abell~2744, we extract source regions with a $1''$ radius centered on each galaxy. The background is measured using local regions around each source to account for the varying foreground ICM level of Abell~2744. We used circular annuli around each source with $3''-6''$ radii  (Figure~\ref{fig:chandra}).

\begin{figure}
\centering
\includegraphics[width=0.9\textwidth]{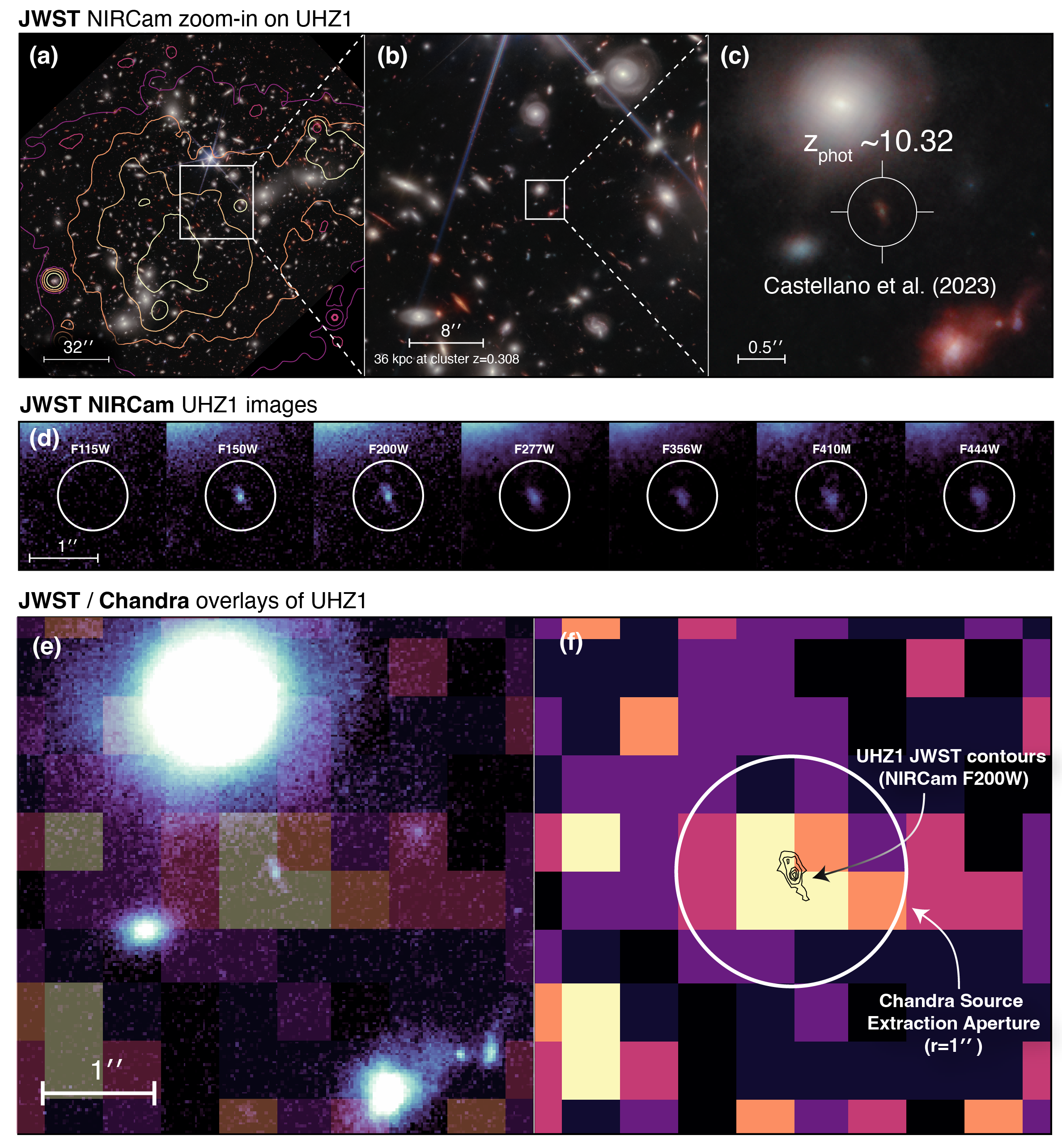}
 \caption{JWST and Chandra images of UHZ1: Panel (a) The \textit{JWST} NIRCam image of the surroundings of UHZ1, and a zoom-in NIRCam image of UHZ1 in Panels (b and c).
 Panel (d) \textit{JWST} images of UHZ1 in seven filters. The galaxy is detected in all \textit{JWST} bands except for  F115W. The non-detection in the bluest F115W band clearly indicates the dropout nature of the galaxy and suggests that it is located at $z\approx10$. The source is extended, with a potentially disturbed morphology evocative of late-stage mergers at lower redshift. A bright nuclear region is apparent in the F150W and F200W bands, and the contrast of this nucleus against the galaxy outskirts decreases for the redder bands. (e) A \textit{JWST}/\textit{Chandra} overlay showing a $4.2\sigma$ excess of X-ray counts cospatial with UHZ1. (f) The same \textit{Chandra} $2-7$ keV \textit{Chandra} image, this time with UHZ1 represented as black contours. The size of the X-ray source is consistent with a point source. The location, luminosity, and spectral characteristics of the source suggest that it is a heavily obscured quasar residing in the $z=10.3$ galaxy, UHZ1. North is up and East is left.}
 \label{fig:chandra_jwst}
\end{figure}

Of this sample of 11 \textit{JWST} galaxies, we detect a statistically significant X-ray source associated with UHZ1 (RA=0:14:16.096, Dec=-30:22:40.285); this galaxy is magnified\cite{castellano22b} by a factor of $\mu=3.81^{+0.41}_{-0.56}$. No other galaxies are located in the vicinity of UHZ1 that could be associated with the X-ray source (Figure~\ref{fig:chandra_jwst}). We note that of the galaxy sample, UHZ1 has the highest lensing magnification factor. Based on the deep \textit{JWST} imaging data, three methods have provided the best fit photometric redshift of $z\approx10.3^{+0.6}_{-1.3}$, owing primarily to its robust non-detection in the F115W band\cite{castellano22b}, as well as further non-detections in \textit{Hubble Space Telescope (HST)} F606W and F814W bands. Moreover, the redshift probability distribution function of UHZ1 does not reveal a secondary peak, making UHZ1 a robust high-redshift galaxy candidate.\cite{castellano22b}. The \textit{Chandra} X-ray source associated with UHZ1 has 42 total and 20.6 net counts within the extraction region in the $2-7$~keV band (rest-frame energy range of $22.6-79.1$~keV). The probability of finding 42 counts with a background expectation of 21.4 counts is $2.73\times10^{-5}$, which is equivalent to a $4.2\sigma$ detection significance assuming a Poisson distribution (Figures~\ref{fig:chandra} and \ref{fig:chandra_jwst}). We emphasize that the particular choice of the background region does not influence the detection of the X-ray source. The source is undetected in the $0.5-2$~keV band (rest-frame energy range of $11.3-22.6$~keV) with the $1\sigma$ upper limit of $<\,2.4$ net counts. The non-detection can be explained by the heavily absorbed nature of the source (see below and the Methods Section). The \textit{Chandra} data of other \textit{JWST}-detected high-redshift galaxies will be discussed in a follow-up study. To further probe the detection significance of the X-ray source, we performed an additional experiment that probed the distribution of the counts in the vicinity of UHZ1. This test, which is described in the Methods Section, suggests that the X-ray source associated with UHZ1 is detected at the $4.4\sigma$ level.

\begin{figure}
\centering
\includegraphics[width=0.9\textwidth]{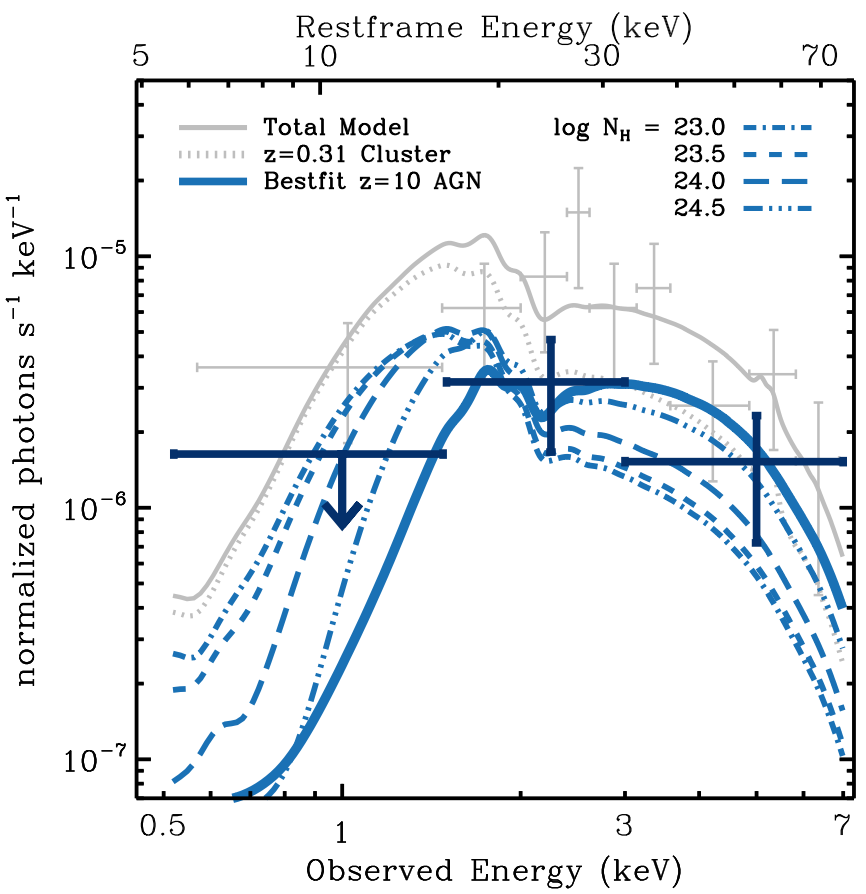}
 \caption{Chandra X-ray spectral energy distribution and model fits: Observed-frame X-ray photons extracted from the merged {\it Chandra} ACIS data in the source aperture containing UHZ1 and the foreground lensing cluster, Abell~2744 (gray crosses; binned to a minimum of 2 counts for plotting purposes). The dotted gray line provides the thermal plasma model (APEC with $kT\sim10.9$~keV) established from a surrounding background annulus and rescaled to match the source aperture area. The solid gray curve provides the best-fit total model of the combined cluster+AGN X-ray emission constructed utilizing the MyTorus library. The observed X-ray spectrum is inconsistent with a pure plasma model - solid blue line shows the best-fit pure AGN model after subtraction of the cluster emission; this model reproduces the background-subtracted broad-band X-ray photometry extracted using forced photometry at the position of UHZ1 ($0.5-1.5$ keV photometric point is a $3\sigma$ upper limit due to the non-detection of X-ray emission after background subtraction). For illustration purposes, we further provide the best-fit pure AGN components for a range of assumed column densities ($N_{\rm H}=$\{0.1,0.3,1.0,3.0\}$\times 10^{24}$~cm$^{-2}$).} 
  \label{fig:spectrum}
\end{figure}

We establish the characteristics of the X-ray source by spectral fitting the \textit{Chandra} X-ray data (Figure~\ref{fig:spectrum}). We extract spectra from the source and background regions, where the latter describes the emission from the ICM that dominates the local background. We fit the background annulus with an optically-thin thermal plasma model (\textsc{apec} in \textsc{XSpec}) with the temperature and normalization being free parameters. We fixed the line-of-sight column density at the Galactic value\cite{column} ($N_{\rm H} = 1.35 \times10^{20} \ \rm{cm^{-2}}$), the metallicity at $0.3\,Z_{\rm \odot}$, and the redshift of Abell~2744 at $z=0.308$. We measure a best-fit temperature for the ICM of $kT=10.9 \pm 1.9$~keV. To fit the spectrum of the source region, we utilize a two-component model. The first component is the \textsc{apec} model established from the background annulus with all parameters fixed and normalization adjusted to the smaller source aperture. The second component describes the heavily absorbed X-ray source and implements a model with a zeroth-order absorbed continuum combined with a Compton-scattered continuum using the MyTorus library\cite{yaqoob12}. We fix the slope of the power law at $\Gamma=1.9$, a disk inclination of 85$^{^\circ}$, and a Compton-scattering fraction of $2\%$, values that are typical of rapidly accreting, heavily absorbed high-redshift AGN\cite{goulding18}.  We find a best-fit ${\rm cstat}=32.5$ with 30 degrees of freedom, which implies an acceptable fit. Further details of the spectral fitting procedure are provided in the Methods Section.  

We obtain a best-fit column density of $N_{\rm H} \approx 8^{+\inf}_{-7} \times 10^{24}$~cm$^{-2}$ and a corresponding intrinsic $2-10$~keV luminosity of $L_{\rm X,int} \approx 9 \times 10^{45}  \ \rm{erg \ s^{-1}} $ after correcting for the $\mu=3.81$ lensing magnification at the location of UHZ1, and accounting for the source extraction region including $\approx85\%$ of the source counts given the \textit{Chandra} point spread function. These results imply that a heavily obscured, most likely Compton-thick, accreting BH is present in UHZ1. Due to the small number of photons, the spectral fitting is somewhat degenerate between $L_{\rm X}$ and $N_{\rm H}$, with larger $N_{\rm H}$ columns requiring significantly larger values of $L_{\rm X}$, and within 2$\sigma$ uncertainties, values as low as $N_{\rm H} > 10^{22}$~cm$^{-2}$ are also allowable within the fit, which would imply a $2\sigma$ lower limit for the $2-10$~keV luminosity of $L_{\rm X,int} > (2-4) \times 10^{43} \ \rm{erg \ s^{-1}} $. 

We use our X-ray spectral measurements to derive the physical properties of the accreting BH in UHZ1. Our best-fit $2-10$~keV luminosity, combined with the appropriate bolometric correction factor\cite{duras20} of $L_{\rm bol}/L_{\rm 2-10keV}\sim 73$, implies a BH mass of $M_{\rm BH} \approx 6 \times 10^9\ \rm{M_{\odot}}$, assuming Eddington-limited accretion. However, this high-mass estimate is driven by (and as noted previously) an $L_{\rm X}$ measurement that is strongly degenerate with $N_{\rm H}$, and is hence, likely unreasonable given the inferred stellar mass of the galaxy. Therefore, we conservatively adopt the lower $1\sigma$ uncertainty measurement for the column density of $N_{\rm H} = 2 \times 10^{24}$~cm$^{-2}$, which provides a sensible balance between the steep degenerate fits required for very high values of $N_{\rm H}$, while still providing a measurement of $L_{\rm X}$ that is allowable across a wide range of potential $N_{\rm H}$. This yields an intrinsic $2-10$~keV luminosity of $L_{\rm X,int} \approx 1.9 \times 10^{44} \ \rm{erg \ s^{-1}} $. Utilizing $L_{\rm bol}/L_{\rm 2-10keV}=21$ at this $L_{\rm X}$ \cite{duras20}, we predict a BH mass of $M_{\rm BH} \approx 4 \times 10^7\ \rm{M_{\odot}}$. Allowing for the quoted factor $\sim$2 uncertainty in the $L_{\rm bol}/L_{\rm 2-10keV}$ relation\cite{duras20}, combined with potential sub-Eddington accretion, we conservatively adopt a mass estimate for the BH in UHZ1 at $z=10.3$ in the range of $10^{7-8}\ \rm{M_{\odot}}.$ Taken together, these combined \textit{JWST} and \textit{Chandra} observations provide unambiguous evidence that UHZ1 harbors a rapidly-accreting and heavily-obscured supermassive BH already in place when the Universe was only 500~Myrs old.

\begin{figure}
\centering
\includegraphics[width=0.9\textwidth]{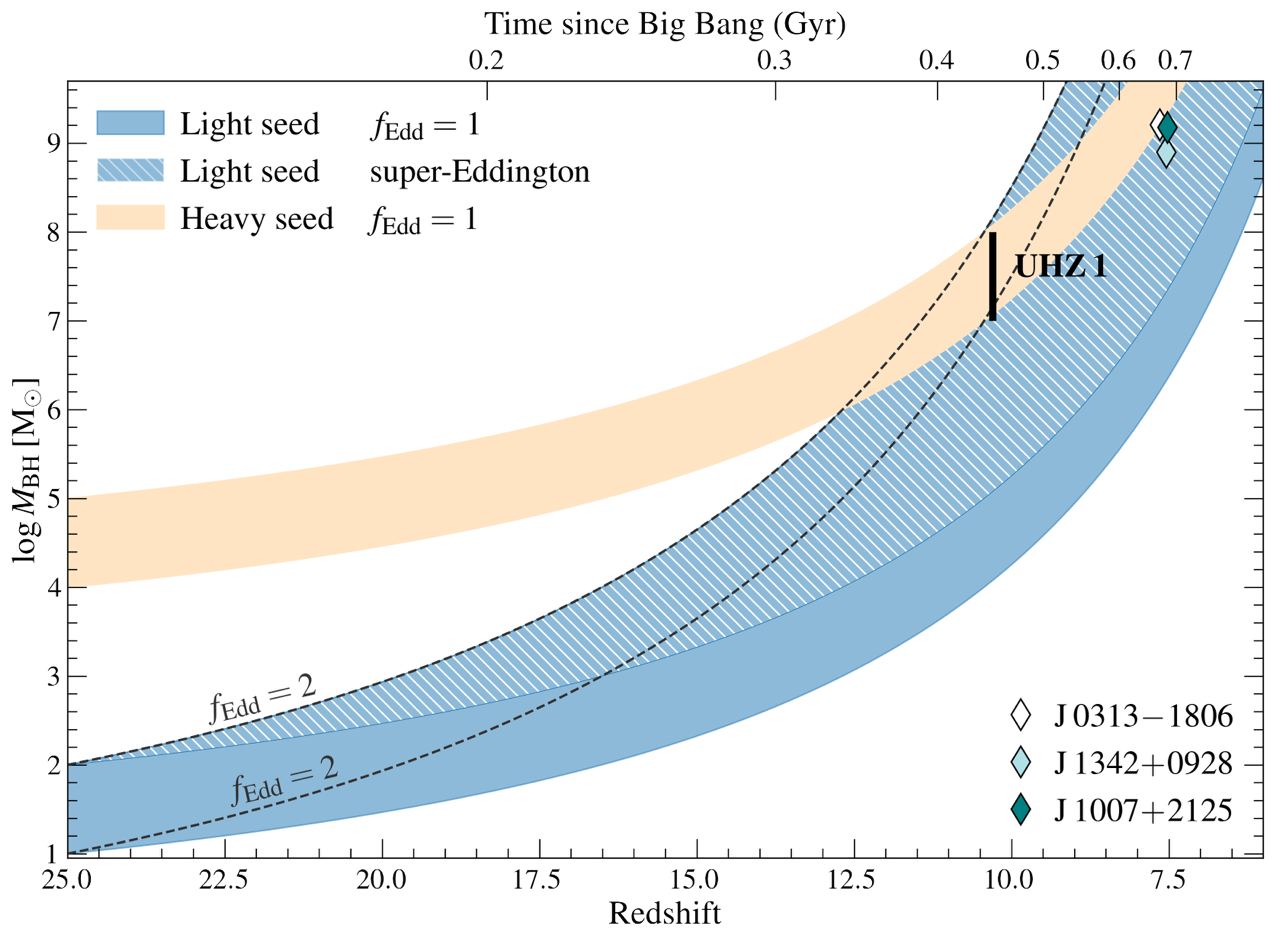}
 \caption{Sketch of the growth of BHs with different initial seed masses and accretion rates: BHs formed via the light seed scenario with $10-100 \,M_{\odot}$ mass can only reach $10^4-10^5 \,M_{\odot}$ by $z=10.3$ if they accrete at their Eddington limit (blue shaded region), which falls short by $2-4$ orders of magnitude of the BH mass estimated for UHZ1. Implausibly high sustained accretion at a rate of at least twice the Eddington limit would be required for light seeds to reach the BH mass close to that of UHZ1 (blue-hatched region). However, for light seeds continuous accretion at the Eddington limit or above for several hundred of million years is highly unlikely as noted by \cite{Willott+10}. Heavy seed models with $10^4-10^5 \,M_{\odot}$ initial BH masses can grow to the mass of the BH powering UHZ1 by $z=10.3$ assuming accretion at the Eddington limit (tan shaded region). All over-plotted models assume a radiative efficiency of $10\%$ and continuous accretion. We also show the location of the three previously known highest redshift quasars at $z\sim7.5$, which were identified in large-area optical surveys\cite{mortlock11,banados18,wang21}. The systematic uncertainty (not shown) on the BH mass of these quasars is $\sim\,0.5$~dex. The mass range shown for UHZ1 corresponds to the derived estimate as noted in the text.}
  \label{fig:bh_growth}
\end{figure}

A range of seeding scenarios operating within the first few hundred million years of the Universe has been proposed to account for the rapid assembly of early BHs, such as the one detected in UHZ1. They can be grouped broadly into ``light seed'' and ``heavy seed'' models. The light seed scenario\cite{madau01} involving the collapse of the first generation - Population III - stars predicts initial BH seeds with $10-100\,M_{\odot}$. The remnant's BH mass estimate suffers from considerable uncertainty stemming from our current poor understanding of the initial mass function (IMF) of Population III stars\cite{HiranoBromm2017,Chon+2022}. Such light seeds are unlikely to grow into massive BHs given their sub-optimal spatial locations and small capture radius\cite{Smith+2018,Pacucci+2017}. Alternately, heavy seed models invoking the formation of massive ($10^{4}\,-\,10^{5}\,M_{\odot}$) BH seeds through the direct collapse of pristine, massive gas clouds or pre-galactic disks\cite{madau01,loeb94,lodato06,begelman06,LodatoPN2007}. While these heavy seeds from direct collapse require rare physical conditions, simulations suggest that they are available in the early Universe\cite{Agarwal+2013,Wise+2019,Lupi+2021}. There have also been additional proposals that suggest formation scenarios for intermediate mass BHs post the formation of Population III stars in the early Universe that blurs this broad dichotomy of light and heavy seeds. After the formation of Population III stars (that will inevitably result in the formation of light seeds), the subsequent episode of star formation is expected to produce dense nuclear star clusters. These environments could, in turn, serve as further new incubators for the formation of intermediate-mass BHs facilitated by stellar and gas dynamical processes that occur in them; from a range of stellar dynamical interactions that can occur in these dense environments\cite{Devecchi+2009} as well as extremely rapid amplified growth of an embedded light seed from wind-fed accretion\cite{AlexanderPN2014,Natarajan2021}. Because information about the initial seeding of BHs is mostly erased during the complex growth and subsequent evolution of BHs over cosmic time\cite{Volonteri+2008,RicartePN2018}, the detection of the X-ray quasar in UHZ1 at $z\approx10.3$ opens up a new, exciting frontier. 

Previous SED fitting of the \textit{JWST} photometric data\cite{castellano22b} suggests that the galaxy has a stellar mass of $M_{\rm \star} = 0.4^{+1.9}_{-0.2} \times 10^8\,M_{\odot}$, which makes it comparable to the inferred BH mass. This implies a strikingly different BH-to-host galaxy stellar mass ratio than observed in the local Universe, where the mass of the central BH is roughly $0.1\%$ of the stellar mass. Such a high BH--galaxy mass ratio is predicted by theoretical studies of high-redshift galaxies seeded with heavy initial BHs\cite{Natarajan+2017}. Heavy seeds and their host galaxies are expected to inevitably transition through such an Outsize Black Hole Galaxy (OBG) stage at early times, before feedback-regulated efficient stellar assembly takes over, eventually leading to the flipping of the mass ratio at later cosmic times\cite{Agarwal+2013,Natarajan+2017}. We note that the current \textit{JWST} SED fitting for UHZ1\cite{castellano22b} assumes that the rest-frame UV/optical emission derives solely from the stellar component modeled with a Salpeter initial mass function (IMF) and that the accreting BH does not contribute appreciably to the SED in the \textit{JWST} bands. By contrast, the rest-frame UV/optical SED for OBGs is predicted to be dominated by accreting BH, as these transient sources are postulated to harbor a growing heavy initial seed with the stellar light contribution from stars modeled with a Population III IMF\cite{Natarajan+2017}. 

According to current BH formation theories, seed BHs may form as early as 200 million years after the Big Bang, implying that the BH in UHZ1 had only $\sim\,300$ million years to grow from its initial seed mass. To interpret the observed properties of the BH powering the X-ray quasar in UHZ1, we trace the mass assembly history of initial BH seeds starting from light and heavy seeds from $z = 25$ to the final inferred BH mass for UHZ1 of $\sim 10^{7-8}\,M_{\odot}$. We note that in this work, the growth history of seeds is tracked from birth to the epoch at which UHZ1 is detected, namely $z=10.3$, and no claims are currently made for the future growth of this source over cosmic time. As shown in Figure~\ref{fig:bh_growth}, an initially light seed with a mass of $10-100\,M_{\odot}$, needs to be consistently accreting at more than twice the Eddington rate throughout; while a heavy seed with an initial mass of $10^{4-5}\,M_{\odot}$ reaches the final mass of the BH powering UHZ1 by accreting at just the Eddington rate. We note from Figure~\ref{fig:bh_growth} that UHZ1, by virtue of its inferred mass and redshift, currently offers more discriminating power viz-a-viz initial seeding models and hence more compelling evidence for heavy initial BH seeds than the three currently known $z\sim7.5$ quasars. Even in the very small regions probed, JWST may now be finding evidence for substantial episodes of BH growth at $z>8$\cite{larson+23,Maiolino+23}. The detection of UHZ1, and potentially other sources, may likely represent only the tip of the iceberg in terms of uncovering the accreting BH population at these early cosmic times. However, as demonstrated in cosmological simulations\cite{DiMatteo+23,DiMatteo+17}, the most massive BH detected at very early epochs does not necessarily go on to produce the highest mass BH at later times as the growth history depends strongly on the details of the environment of these sources. Therefore, these early sources like UHZ1 may not necessarily all be the progenitors for the observed optically-detected quasars harboring $10^{9-10}\,M_{\odot}$ BHs at $z \sim 6-7$. With additional data that is forthcoming and dedicated follow-up studies, with a better census, abundance estimates will soon be possible.

For a light initial BH seed in UHZ1, regular accretion at more than twice the Eddington rate for a period of 300 million years requires gas to be continuously delivered to the nucleus after efficient removal of its angular momentum. A gas reservoir in excess of $10^7-10^8\,M_{\odot}$ would be needed to continuously feed the BH at this rate for several hundred million years. Numerical studies\cite{Alvarez+2009,Kim+2011}, including the state-of-the-art Renaissance simulation suite, demonstrate that this is likely infeasible given the spatial location of light seeds given their Population III origin. By tracing the growth history of $\sim\,15,000$ light seeds, across three differing over-density zoom-in regions with parsec scale resolution, populated by mini-haloes ranging in mass from $10^6-10^9\,\rm{M_{\odot}}$, no significant growth by accretion was found. Even the most active BHs grew in mass by at most 10\% over the seed mass across several hundred million years.\cite{Smith+2018}. Though many mini-haloes contained dense, cool gas clumps to accrete from, light seeds rarely inhabited such regions. In fact, star formation was found to compete with BH accretion in terms of gas consumption; and the resultant feedback processes in turn effectively suppressed BH growth. Furthermore, as light seeds are predicted to be produced from supernovae explosions of Population III stars, they are typically born into evacuated, low gas density environments, bereft of gas, potentially leading to their stunted growth\cite{Smith+2018}.

While super-Eddington accretion is ubiquitously realized in idealized General Relativistic Magneto Hydro-dynamics simulations, these simulations do not adopt cosmological initial conditions, since their domain of integration extends only to $\sim\,1000$ gravitational radii; furthermore, they do not track the evolution of BHs on timescales of several hundred million years\cite{Jiang+2019}. Additionally, cosmological simulations indicate that the feedback resulting from super-Eddington accretion rates curtails BH growth rather than amplifies it, in a variety of astrophysical environments, including the birth site of heavy seeds\cite{Regan+2019,Sassano+2023,Massonneau+22}. Therefore, starting with an initially light seed to reach the inferred BH mass of UHZ1 by $z\approx10.3$ requires multiple finely tuned conditions that have been demonstrated to be implausible in current cosmological simulations.  Additionally, physical conditions at $z=10.3$ are markedly different, as the Universe has not yet reionized, in comparison to $z \sim 6-7$. Therefore, studies of the sizes of proximity zones of quasars studied at those later epochs \cite{Eilers+2017} that have been used to make a case for episodic super-Eddington accretion for a small fraction of the population may not capture the conditions at $z=10.3$. Besides, simulations tailored to study high-redshift quasar absorption spectra also find that proximity zones sizes are highly epoch-dependent and compactness could result from the presence of Damped Lyman-alpha or Lyman Limit systems along the line of sight in the IGM rather than telegraph information about accretion physics \cite{ChenGnedin2021}. Meanwhile, as we demonstrate, an initially heavy seed growing even modestly at the Eddington rate or even slightly below can comfortably reach the inferred mass of UHZ1.

\newpage

\begin{center}
\begin{LARGE}\textbf{Methods}\end{LARGE}
\end{center}

\vspace{0.5cm}
\noindent
\begin{large}\textbf{The Chandra data}\end{large}
\vspace{0.3cm}

The high spatial resolution \textit{Chandra} data of Abell~2744 is the backbone of the analysis presented in this work. The processing and analysis of the data were carried out with standard \textsc{CIAO}\cite{fruscione06} tools. In this work, we used 60 individual \textit{Chandra} observations that are listed in Table~1. The total exposure time of all observations is 1.25~Ms. The first step of the analysis was to reprocess the individual observations using the \textsc{chandra\_repro} tool to apply the latest calibration data. Observations with \textit{Chandra}'s ACIS-I detector are not prone to background flares, which was verified for each Abell~2744 observation. Although \textit{Chandra}'s absolute pointing accuracy is better than $0.4''$, we implemented the \textsc{wcs\_match} and \textsc{wcs\_update} tools to further improve the accuracy of the astrometry. To this end, we created a frame transformation for each observation by utilizing a set of bright X-ray sources in the Abell~2744 field, thereby minimizing the aspect difference. The aspect-corrected observations were co-added with the \textsc{merge\_obs} task using ObsID 8477 as the reference coordinate system. With this tool, we generated merged images in the $0.5-7$~keV, $0.5-2$~keV, and $2-7$~keV energy ranges. Similarly, we created maps of the point spread function (PSF) for each observation and each energy range, which were then weighted, and co-added. The merged images were used to carry out the X-ray photometry. To probe the \textit{Chandra} PSF at the location of the UHZ1 X-ray source, we utilized the MARX ray tracing simulation suite (https://space.mit.edu/cxc/marx/). We applied the \textsc{simulate\_psf} tool and simulated the PSF for each observation assuming the $E=2.5$~keV peak energy of the photons. Based on the co-added simulated PSF maps, we found that an $r=1''$ circular region encircles $\approx88\%$ of the source counts at the location of UHZ1. To probe the distribution of the X-ray photons associated with UHZ1, we extracted the number of source and background counts as a function of radius from $r=0.5''$ to $r=1.5''$ and derived the distribution of net (i.e.\ background subtracted) counts. We found that the distribution of X-ray photons associated with UHZ1 is in excellent agreement with the PSF model, which confirms that the X-ray source in UHZ1 is a point source and that the $r=1''$ region includes $\approx88\%$ of the source counts.

As discussed in the main body of the paper, we carry out the X-ray photometry using a $1''$ circular region centered on UHZ1\cite{castellano22b}. The applied background region is a circular annulus with $3''$ and $6''$ inner and outer radii, respectively. The X-ray photometry was performed using the \textsc{dmextract} tool. We detected 42 counts from the source region with a background expectation of 21.4 counts, implying the presence of 20.6 net counts. Given the underlying and spatially-variable emission from the intracluster medium (ICM) of Abell~2744, applying a local background region is most suitable. Because spatial variations in the level of the ICM emission around UHZ1 may influence the observed detection significance, we probed the stability of the ICM emission in the vicinity of the galaxy. Therefore, we measured the $2-7$~keV band X-ray surface brightness level of the emission using six circular annuli with a width of $2''$ between $2''$ to $14''$ radii. We found that the X-ray surface brightness of the emission is fairly uniform and its level exhibits $<3\%$ variations in these annuli, which is comparable with the statistical uncertainties. Given the low-level spatial variation, we conclude that our particular choice of the background region does not influence our ability to detect the X-ray source co-spatial with UHZ1. Additionally, we also investigated the statistical uncertainty that stems from the finite number of counts in the background. Given the number of background counts in the $3''-6''$ annulus, we derive that the uncertainty associated with Poisson statistics is $\approx4.4\%$. Given the background expectation of 21.6 counts in the source region, this implies an uncertainty of $0.95$ counts.     

Because the aim of this work is to search for X-ray sources associated with \textit{JWST}-detected galaxies, we probed the astrometric consistency between \textit{Chandra} and \textit{JWST}. We identified a sample of bright X-ray sources with \textit{JWST} counterparts and derived the offset between the X-ray and near-infrared centroids. Based on the individual offsets, we computed the average shift and the dispersion in the Abell~2744 field. We found that the average offset is $\delta = 0.33''$ with a dispersion of rms = 0.14, less than the \textit{Chandra} pixel size. The small offset highlights the consistency between the \textit{Chandra} and \textit{JWST}  astrometry and demonstrates that inaccuracies in the absolute astrometry do not present a systematic uncertainty in the detection of the X-ray source at the position of UHZ1. 

\begin{sidewaystable}
\begin{center}
\caption{List of analyzed \textit{Chandra} observations}
\begin{minipage}{15cm}
\renewcommand{\arraystretch}{1}
\centering
\begin{tabular}{ cccc|cccc}
\hline
Obs.\,ID & $t_{\rm exp}\ \rm{(ks)}$ & Detector & Obs.\,Date & Obs.\,ID & $t_{\rm exp} \ \rm{(ks)}$ & Detector & Obs.\,Date \\
\hline
2212     &    24.82    &    ACIS-S    &    2001-09-03     &   25938    &    18.66   &    ACIS-I   &    2022-11-26  \\  
7712     &    8.07     &    ACIS-I    &    2007-09-10     &   25939    &    14.32   &    ACIS-I   &    2023-01-28  \\  
7915     &    18.62    &    ACIS-I    &    2006-11-08     &   25942    &    15.18   &    ACIS-I   &    2022-05-04  \\ 
8477     &    45.91    &    ACIS-I    &    2007-06-10     &   25944    &    21.62   &    ACIS-I   &    2022-09-08  \\ 
8557     &    27.81    &    ACIS-I    &    2007-06-14     &   25945    &    17.02    &   ACIS-I   &   2022-09-27   \\
25278    &    9.78     &    ACIS-I    &    2022-12-02     &   25948    &    27.87    &   ACIS-I   &   2022-09-30   \\
25279    &    24.46    &    ACIS-I    &    2022-09-06     &   25951    &    28.71   &    ACIS-I   &    2022-11-18  \\
25907    &    36.80    &    ACIS-I    &    2022-11-08     &   25953    &    24.75   &    ACIS-I   &    2022-09-17  \\
25908    &    22.61    &    ACIS-I    &    2022-09-23     &   25954    &    13.40    &   ACIS-I   &   2022-04-24   \\
25910    &    19.31    &    ACIS-I    &    2022-09-25     &   25956    &    13.90   &    ACIS-I   &    2022-09-02  \\
25911    &    16.85    &    ACIS-I    &    2022-04-19     &   25957    &    21.80   &    ACIS-I   &    2022-09-08  \\
25912    &    15.37    &    ACIS-I    &    2022-04-18     &   25958    &    12.32   &    ACIS-I   &    2022-05-04  \\
25913    &    19.64    &    ACIS-I    &    2022-09-03     &   25963    &    37.59   &    ACIS-I   &    2022-11-26  \\
25914    &    28.20    &    ACIS-I    &    2022-10-15     &   25967    &    33.64   &    ACIS-I   &    2022-08-01  \\
25915    &    21.08    &    ACIS-I    &    2022-09-03     &   25968    &    27.46   &    ACIS-I   &    2022-07-12  \\
25918    &    20.63    &    ACIS-I    &    2022-09-13     &   25969    &    27.72   &    ACIS-I   &    2022-10-09  \\
25919    &    25.29    &    ACIS-I    &    2022-06-13     &   25970    &    24.75   &    ACIS-I   &    2022-06-12  \\
25920    &    30.51    &    ACIS-I    &    2022-06-13     &   25971    &    12.61   &    ACIS-I   &    2022-05-04  \\
25922    &    31.37    &    ACIS-I    &    2022-06-14     &   25972    &    31.75   &    ACIS-I   &    2022-05-18  \\
25923    &    10.94    &    ACIS-I    &    2022-09-04     &   25973    &    18.15   &    ACIS-I   &    2022-11-11  \\
25924    &    21.79    &    ACIS-I    &    2022-09-07     &   26280    &    11.74   &    ACIS-I   &    2022-01-18  \\
25925    &    23.60    &    ACIS-I    &    2022-09-02     &   27347    &    21.95   &    ACIS-I   &    2022-09-09  \\
25928    &    15.87    &    ACIS-I    &    2022-05-03     &   27449    &     9.78   &    ACIS-I   &    2022-09-24  \\
25929    &    27.72    &    ACIS-I    &    2022-08-26     &   27450    &     9.78   &    ACIS-I   &    2022-09-26  \\
25930    &    19.92    &    ACIS-I    &    2022-11-15     &   27556    &    25.15   &    ACIS-I   &    2022-11-15  \\
25931    &    14.58    &    ACIS-I    &    2022-04-23     &   27575    &    19.65   &    ACIS-I   &    2022-12-02  \\
25932    &    14.08    &    ACIS-I    &    2022-05-05     &   27678    &    12.42   &    ACIS-I   &    2023-01-27  \\
25934    &    19.30    &    ACIS-I    &    2022-04-21     &   27679    &    11.93   &    ACIS-I   &    2023-01-28  \\
25936    &    12.92    &    ACIS-I    &    2023-01-26     &   27680    &    13.21   &    ACIS-I   &    2023-01-28  \\
25937    &    30.78    &    ACIS-I    &    2022-11-27     &   27681    &     9.78   &    ACIS-I   &    2023-01-29  \\
 \hline
\label{tab:data}
\end{tabular} 
\end{minipage}
\end{center}

\end{sidewaystable}

\vspace{0.5cm}
\noindent
\begin{large}\textbf{Spectral fitting of the X-ray data}\end{large}
\vspace{0.3cm}

X-ray counts were extracted from the merged Level 2 events file using the CIAO package \textsc{specextract} along with associated response matrices. Two apertures were used for the photon extraction, 1) an $r=0.8''$ circular aperture at the position of UHZ1, and 2) an annulus with an inner radius of $3''$ and outer radius of $6''$ centered at UHZ1. The annulus region represents a ``clean'' region containing only emission arising from the foreground cluster gas, while the smaller source region contains emission from both the foreground cluster and UHZ1. Using the \textsc{grppha} package, we grouped the extracted photons to contain at least one count per spectral bin for the use of Cash statistics during the fitting process\cite{cash79}.

\begin{figure}
\centering
\includegraphics[width=0.9\textwidth]{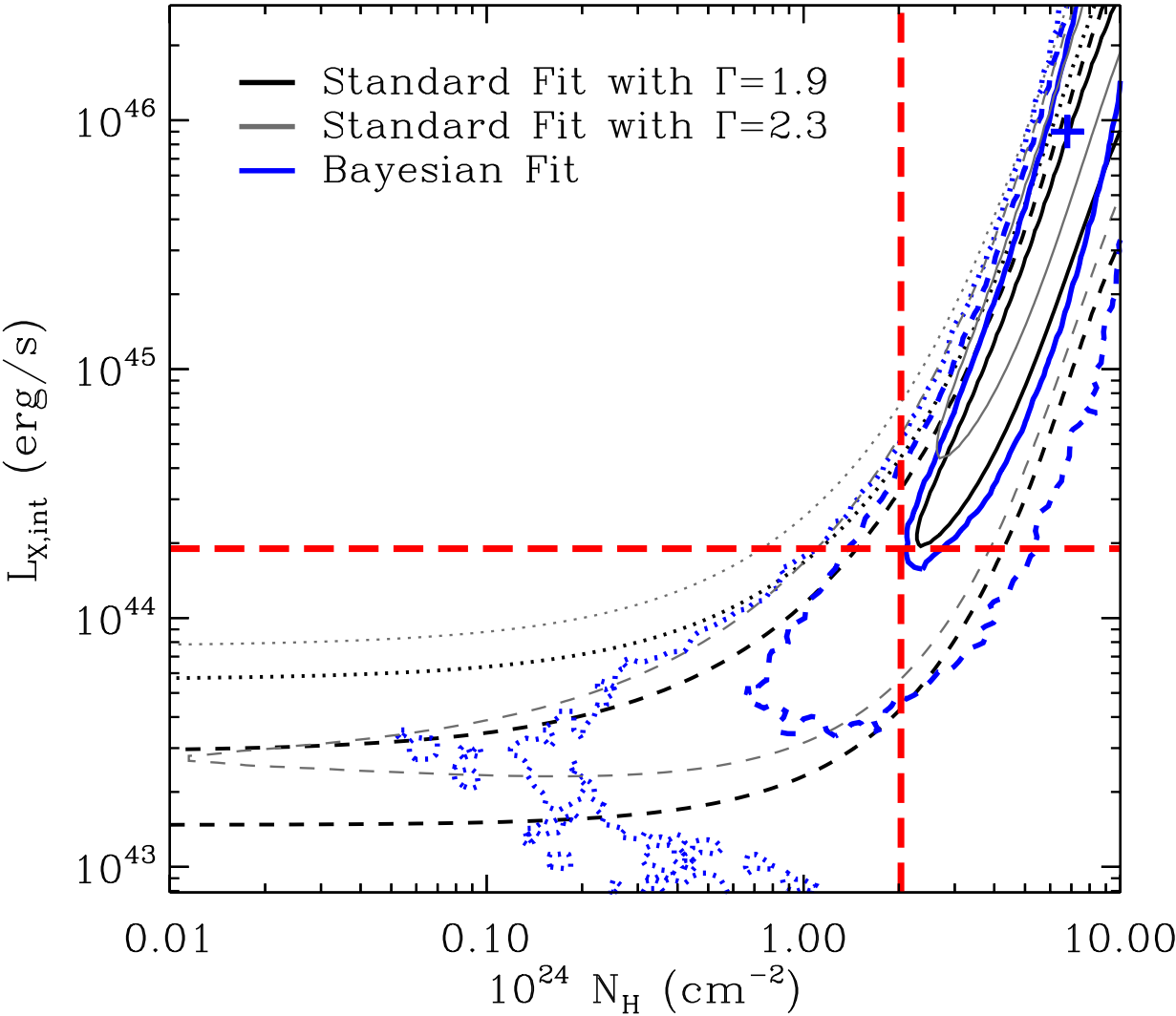}
 \caption{Best-fit statistical uncertainties and posterior draws of the demagnified intrinsic X-ray luminosity ($L_{\rm X,int}$) and gas column density ($N_{\rm H}$) of UHZ1. One, 2, and 3$\sigma$ (solid, dashed, and dotted lines, respectively) contours are constructed from the X-ray spectral fits to the {\it Chandra} data. Black lines show the statistical fits for a fixed AGN spectral slope of $\Gamma=1.9$ and assuming fixed best-fit (rescaled) normalization and plasma temperature for the background APEC component. Light-gray color represents a fixed slope of $\Gamma=2.3$. Blue lines are the posterior draws from a Bayesian fit employing broad informative Gaussian priors for $\Gamma$, $n_{\rm APEC}$ and kT. The mode of the posterior is shown with a blue cross. For values of $N_{\rm H} \gtrsim 2\times 10^{24}$ (equivalent to the lower $1\sigma$ contours), a strong steep degeneracy between $L_{\rm X,int}$ and $N_{\rm H}$ is clearly evident, we thus adopt this threshold to mitigate these degeneracies, resulting in an adopted $L_{\rm X} \sim 1.9 \times 10^{44}$~erg/s.}
 \label{fig:Xray_contours}
\end{figure}

We proceed by first fitting the cluster-only X-ray emission using \textsc{XSPEC} version 12.13 assuming an \textsc{APEC} model attenuated by an additional photoelectric absorber arising from material in the Milky Way with $N_{\rm H} = 1.35 \times 10^{20}$~cm$^{-2}$\cite{column}. We froze the previously known redshift of the cluster to $z=0.308$ and set an abundance of 0.3 Solar\cite{leccardi08}. We find a best-fit temperature of $kT=10.9 \pm 1.9$~keV and normalization of $(17.23 \pm 0.57) \times 10^{-6}$ with C-stat = 277.7 (for 323 degrees of freedom). This fit is consistent with temperature measurements of the intracluster medium around UHZ1\cite{owers11}. We use this best-fit cluster-gas model as a fixed and known background in the smaller circular source aperture after renormalizing to account for the differences in aperture sizes. Employing this pre-determined model directly to the photons extracted from the source aperture we find a poor fit with cstat=42.2 and 32 degrees of freedom. We note that if we do not assume a known value, and instead fit for the gas temperature, the required temperature is pegged at the maximally allowed temperature ($kT\sim 67$~keV) for the APEC model, owing to the presence of additional high energy photons in the source aperture. 

We next fit the source aperture using a combination of the pre-determined aperture-corrected cluster emission and a redshifted absorbed power-law component to represent a distant point source. The redshift of the model was set to $z=9.99$ i.e., the photometric redshift solution determined of UHZ1\cite{castellano22b} using the updated EaZY spectral template library specifically designed for use with \textit{JWST} data\cite{larson22}. However, we found that a simple redshifted photoelectric absorber model (\textsc{zphabs}) requires an extremely high column density in excess of $N_{\rm} > 10^{25}$~cm$^{-2}$, and thus we instead opt to use the MyTorus library\cite{yaqoob12} to model the absorption as these models are better suited to the complex emission arising in high-$N_{\rm H}$ environments. We assume a relatively minimal model with a zeroth-order absorbed continuum combined with a Compton-scattered continuum. Given the low-photon statistics, we set a power-law slope of $\Gamma=1.9$, a disk inclination of 85$^{^\circ}$, and a Compton-scattering fraction of $2\%$, all of which are typical of rapidly accreting heavily absorbed AGN\cite{goulding18}. We find a best-fit cstat=32.5 with 30 degrees of freedom, i.e., a $\Delta$C-stat = 9.7 over the \textsc{APEC}-only model. For the two additional degrees of freedom required, the APEC$+$AGN model passes an F-test at the $>99\%$ significance level over the simple APEC-only model and thus provides very strong evidence for the presence of an additional X-ray point source. 

The best-fit column density is found to be $N_{\rm H} \approx 8^{+\inf}_{-7} \times 10^{24}$~cm$^{-2}$, with an intrinsic $2-10$~keV luminosity of $L_{\rm X,int} \approx9 \times 10^{45}  \ \rm{erg \ s^{-1}} $ (using the magnification factor of $\mu = 3.81$ established for UHZ1\cite{castellano22b}). However, both of these measurements are degenerate in the current data, particularly in the $N_{\rm H} > 10^{24}$~cm$^{-2}$ regime (see Figure~\ref{fig:Xray_contours}). An analysis of the C-stat space shows that within 2$\sigma$, values as low as $N_{\rm H} = 10^{22}$~cm$^{-2}$ are allowable, owing mainly to the diminished soft-energy response of Chandra's ACIS detector. Moreover, we determine that due to the high-energy restframe coverage, for column densities of $N_{\rm} < 10^{24}$~cm$^{-2}$, the allowable range of $L_{\rm X,int}$ is very stable, $\sim (2-4) \times 10^{43} \ \rm{erg \ s^{-1}}$, which we consider as the lower limit on the X-ray luminosity. 

We determined that the precise parameterization of the presumed AGN component has a mild effect on our subsequent $L_{\rm X}$ and $N_{\rm H}$ measurements. For example, we tested the effect of our assumption of $\Gamma = 1.9$ on our derived values. From X-ray analysis of $z\sim6-7$ quasars, a marginal steepening of the spectral slope to $\Gamma \sim 2.3$ was found \cite{wang21}. Assuming this steeper value, we find no significant change in our measured $N_{\rm H}$, but find a systematic shift to larger $L_{\rm X,int}$ by a factor $\sim$1.4. Thus, within our uncertainties, we find no significant difference. Furthermore, significantly larger values for the scattering fraction are ruled out by the lack of detected soft photons. Similarly, lower inclination angles $i<75$~degrees are also disfavored due to insufficient soft emission. Moreover, the precise values of disk inclinations in the range $i = 75$--90~degrees have a weak systematic effect on $N_{\rm H}$, but these variations are well within our uncertainties. 

We further note that fixing the parameters of the APEC model, used to parameterize the foreground cluster emission, to the best-fit values found during the background analysis may also have an effect on the measurements of the AGN component. In an attempt to simultaneously fold in all of these uncertainties, we opted for an additional Bayesian treatment of the data. We applied physically-motivated informative priors on the AGN spectral slope parameterized as $P(\Gamma)\sim N(1.9; 0.15)$, the APEC normalization with $P(n_{\rm APEC})\sim N(3.5\times10^{-7}; 0.2\times10^{-7})$ and plasma temperature with $P(kT)\sim N (10.9; 1.9)$. These priors were designed to reflect the previous knowledge of AGN spectra as well as the highly constrained fit to the background spectra. By contrast, flat uninformative priors were applied to both $L_{\rm AGN}$ and $N_{\rm H}$. We performed a Markov-Chain Monte-Carlo using a Goodman-Weare algorithm with 10 walkers, $10^6$ draws, and a burn-in of $10^5$ per chain. We find that the posteriors of the draws for the APEC component are dominated by the priors, as is expected. We note that using an uninformative set of priors on the APEC component tends to a luminosity that is a factor $\sim 3$ greater than the background expectation and to an unfeasible temperature of $kT > 60$~keV. The joint posterior distribution for $N_{\rm H}$ and $L_{\rm AGN}$ are provided in Figure~\ref{fig:Xray_contours}, and show remarkable consistency with our previous and more simple frequentist fit, albeit now with marginally larger 1-$\sigma$ uncertainties around the peak of the posterior driven by the non-fixed nature of $\Gamma$ and the background APEC model. The lower-1$\sigma$ values for $N_{\rm H}$ and $L_{\rm X}$ are consistent between the Bayesian and frequentist fits, and show the same steep degeneracy between $N_{\rm H}$ and $L_{\rm X}$ towards higher values. Interestingly, we find that the allowable range of $N_{\rm H}$ is more constrained than found previously, owing to the ability of the AGN spectral slope to becoming marginally flatter, as well as the allowance of a cooler plasma temperature, to balance the soft-X-rays. The Bayesian analysis places a 95\% lower limit on the column density of $N_{\rm H} \sim 7 \times 10^{23}$~cm$^{-2}$, and $L_{\rm X} \sim 4 \times 10^{43}\ \rm{erg \ s^{-1}}$, which corresponds to a BH mass of $M_{BH} \sim 6\times 10^6 \,M_{\odot}$ assuming accretion at the Eddington limit. Because BHs formed via light seeds can only reach $10^4-10^5 \,M_{\odot}$ by $z=10.3$ if they accrete at their Eddington limit, the overall conclusions presented in the main body of the paper remain the same even if we assume the $95\%$ lower limit on the X-ray luminosity. Finally, we note that significantly more sensitive data taken with the next generation of X-ray telescopes will be required to provide a more in-depth analysis of the X-ray AGN properties of UHZ1.

\vspace{0.5cm}
\noindent
\begin{large}\textbf{Verifying the statistical significance of the detection}\end{large}
\vspace{0.3cm}

\begin{figure}
\includegraphics[width=0.9\textwidth]{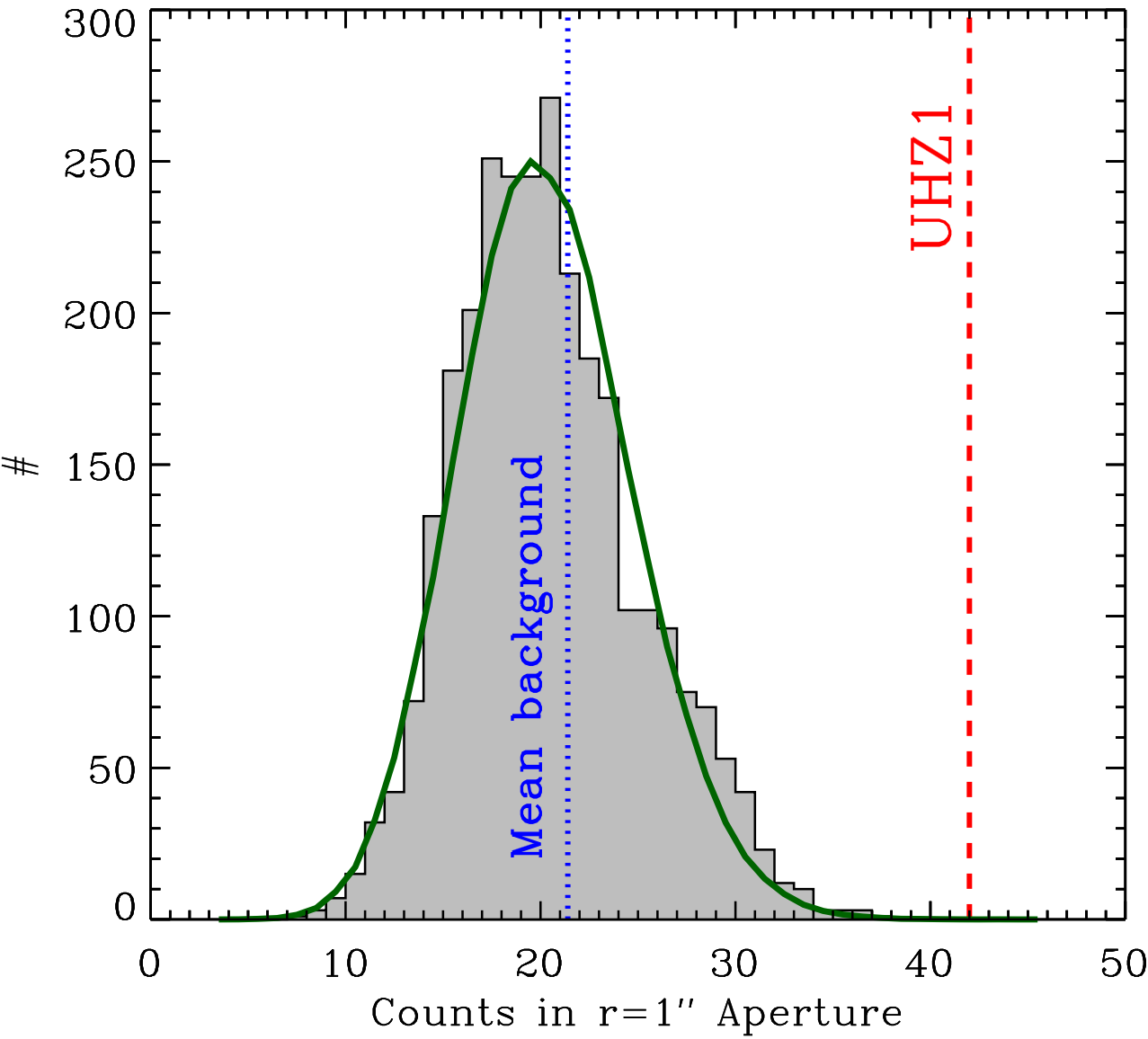}
 \caption{Distribution of background counts extracted from individual $r=1''$ apertures in a $25'' \times 25''$ box centered on the position of UHZ1. The central $r=2''$ region is masked to exclude contaminating X-ray photons from UHZ1 itself. The mean Gaussian background ($\sim 21.4$ photons; blue dotted line) is found to be in good agreement with the best-fit Poisson mean ($\sim 19.9$). The best-fit Poisson distribution shows that the probability of detecting a point source with 42 counts (red dash line) is $1.05 \times 10^{-5}$, and hence, we conclude that UHZ1 is robustly detected with a $4.2-4.4\sigma$ significance.}
 \label{fig:Xray_background}
\end{figure}

Using the $r=1''$ source extraction region we found that the X-ray source associated with UHZ1 is detected at the $4.2\sigma$ statistical confidence level. To further probe the detection significance, we performed an additional statistical test. From the $2-7$~keV band \textit{Chandra} X-ray image, we cut out a $25'' \times 25''$ region. We construct a background map within this rectangular region, by making the ansatz that a point source with $r=1''$ radius exists at the location of each given pixel in the map. To avoid contamination by the bright source associated with UHZ1, we masked an $r=2''$ region around this X-ray source. Given the size of the selected region, there are $\approx2500$ potential aperture positions. We measured the number of total counts associated with each region and present the distribution of the counts in Figure~\ref{fig:Xray_background}. Outside of the masked region containing UHZ1, none of the regions include 42 counts (i.e., the detected number of photons for the X-ray source in UHZ1). The best-fit Poisson mean of the count distribution is 19.9 counts, which is marginally lower than the Gaussian background expectation for UHZ1 ($\sim 21.4$). Based on the best-fit, we conclude that the probability of observing 42 or more counts from a Poisson distribution with $\mu=19.9$ is $P(X\geq42) = 1.05 \times 10^{-5}$, which is equivalent to a $4.4\sigma$ detection. This result is in agreement with the value derived in the main body of the paper, where we obtained a $4.2\sigma$ detection significance for UHZ1 using standard aperture photometry methods. Thus, this independent analysis quantitatively and statistically confirms the robust nature of the X-ray detection.

\vspace{0.5cm}
\noindent
\begin{large}\textbf{Star-formation and clumping cannot be the origin of X-ray emission associated with UHZ1}\end{large}
\vspace{0.3cm}

Although the characteristics of the X-ray source associated with UHZ1 are consistent with those of a heavily obscured AGN, we explore if the source may have a different origin. We explore two possibilities: (1) star formation associated with UHZ1; (2) gas clumping in the intracluster medium of Abell~2744.

In actively star-forming galaxies the total X-ray luminosity originating from X-ray binaries is proportional to the star-formation rate of galaxies. This relationship is robust and has been calibrated for local\cite{mineo12} and for distant galaxies\cite{mineo14,lehmer16}. The observed $L_{\rm X} - \rm{SFR}$ relation can be described with a linear scaling relation and it shows an increasing trend with higher redshift due to the lower metallicity of galaxies. Using the redshift-dependent $L_{\rm X} - \rm{SFR}$ scaling relation\cite{lehmer16} and the star formation rate of UHZ1 ($\rm{SFR}=4.4 \ \rm{M_{\odot} \ yr^{-1}}$), we predict a $2-10$~keV band X-ray luminosity of $\sim4\times10^{41} \ \rm{erg \ s^{-1}}$. This value falls substantially short of the observed X-ray luminosity by about $2-4$ orders of magnitude from the measured X-ray luminosity of UHZ1. As a caveat, we note that the redshift dependence of the $L_{\rm X} - \rm{SFR}$ relation has only been tested to $z\sim4$, implying that a steeper redshift dependence is feasible for galaxies in the early universe. However, an $L_{\rm X} - \rm{SFR}$ relation that is two orders of magnitude higher than this relation would be inconsistent with stacking observations of the Chandra Deep Field South\cite{vito22} and the RELICS clusters\cite{bogdan22}.

We further investigated the possibility that the excess X-ray emission is associated with the intracluster medium of Abell~2744, and is potentially in projection with UHZ1. In this scenario, the emission does not originate from a distant point source, but from a small, dense clump of gas in the intracluster medium. However, dense gas clumps are predicted to be cooler than the surrounding intracluster medium\cite{vazza13}. This, however, is inconsistent with the observed hard energy spectrum extracted from the source aperture. For example, for a gas temperature of $ kT\sim 5$~keV (roughly half of that of the observed temperature of the intracluster medium), we would expect to detect a comparable number of counts in the $0.5-2$~keV and $2-7$~keV bands. This is inconsistent with the \textit{Chandra} observations, as the source is robustly undetected below $E < 2$~keV. Indeed by contrast, and as discussed in the previous section, the observed X-ray spectrum requires a substantially hotter (kT$\sim$67~keV) temperature in order to fit with a simple thermal plasma, not cooler. Thus, we conclude that the excess X-ray emission cannot be associated with the intracluster medium of Abell~2744. \\

\smallskip

\begin{small}

\noindent
\textit{Data Availability:} The \textit{JWST} data of Abell~2744 is publicly available at MAST (http://archive.stsci.edu). The \textit{Chandra} data is available upon request. \\

\noindent
\textit{Author contributions:} \'A.B. is the Principal Investigator of the \textit{Chandra} proposal, analyzed the \textit{Chandra} observations, led the analysis and drafted the paper. A.D.G and P.N. contributed equally. A.D.G. carried out the spectral fitting of the X-ray source, played a major role in writing the manuscript, and provided figures. P.N. led the theoretical interpretation and played a major role in writing the manuscript. O.E.K. contributed to the analysis of the Chandra data and its interpretation and provided a figure. G.R.T. contributed figures and text to the manuscript. U.C. contributed to the sample selection and data analysis. M.V. contributed to the interpretation and text of the manuscript. R.P.K. contributed to the interpretation of the paper. W.R.F., C.J., E.C., and I.Z. reviewed the manuscript and contributed to the text. \\

\noindent 
\textit{Acknowledgements:} We thank the anonymous reviewers for their constructive reports that have allowed us to clarify and improve several aspects of the manuscript. This research has made use of data obtained from the Chandra Data Archive and software provided by the Chandra X-ray Center (CXC) in the application packages CIAO and Sherpa. The authors thank helpful discussions with Fabio Pacucci, Angelo Ricarte, and Peter Edmonds. \'A.B., G.R.T., R.P.K., C.J., and W.R.F. acknowledge support from the Smithsonian Institution and the Chandra X-ray Center through NASA contract NAS8-03060. A.D.G. acknowledges support from NSF/AAG grant No. 1007094. P.N. acknowledges support from the Black Hole Initiative at Harvard University, which is funded by grants from the John Templeton Foundation and the Gordon and Betty Moore Foundation. O.E.K. is supported by the GA\v{C}R EXPRO grant No. 21-13491X. \\

\noindent
\textit{Competing Interests:} The authors declare that they have no competing financial interests. \\

\noindent
\textit{Correspondence:} Correspondence and requests for materials should be addressed to \'AB (email: abogdan@cfa.harvard.edu).

\end{small}

\bibliographystyle{naturemag}
\bibliography{paper1.bib} 


\end{document}